\documentclass[runningheads]{llncs}
\usepackage[T1]{fontenc}
\usepackage{url}
\usepackage{tcolorbox}
\usepackage{amssymb} 
\usepackage{longtable}
\usepackage{xcolor}
\usepackage{graphicx}
\begin{document}

\newcommand{\nb}[2]{
	\fbox{\bfseries\sffamily\scriptsize#1}
	{\sf\small$\blacktriangleright$\textit{#2}$\blacktriangleleft$}
}
\newcommand\TS[1]{\textcolor{blue}{\nb{Tiziano}{#1}}}
\newcommand\MM[1]{\textcolor{green}{\nb{Mahyar}{#1}}}
\newcommand\FD[1]{\textcolor{red}{\nb{Francesco}{#1}}}
\title{SAKE: Software Architectural Knowledge Evaluation Benchmark for Large Language Models}

\titlerunning{SAKE: Software Architectural Knowledge Evaluation Benchmark}

\author{Tiziano Santilli\orcidID{0009-0005-9505-3846}\textsuperscript{*}\and
  Francesco Daghero\orcidID{0000-0001-9595-7216}\textsuperscript{*}\and
Mayhar Tourchi Moghaddam\orcidID{0000-0001-5028-7546}}
\authorrunning{T. Santilli et al.}
\institute{University of Southern Denmark, Odense 5300, DK \\
\email{\{tisa,fdag,mtmo\}@mmmi.sdu.dk} \\
\textsuperscript{*}These authors contributed equally to this work.}
\maketitle              
\begin{abstract}
Large Language Models (LLMs) are increasingly used as assistants across the software
development lifecycle, yet their ability to reason about software architecture remains
largely unmeasured. Architectural decision-making depends on quality attribute trade-offs,
design patterns, and system-level constraints, none of which are exercised by benchmarks
that target syntactic or algorithmic tasks. We introduce SAKE (Software Architectural
Knowledge Evaluation), a standardized and reproducible benchmark for assessing software
architectural knowledge in LLMs. SAKE comprises 2154 expert-curated multiple-choice
questions, each with four options, stratified across eight architectural categories and four
context-length levels. We evaluate 11 proprietary and open-weight models in zero-shot and
five-shot settings. Overall accuracy is high, but performance varies markedly across
categories, revealing competency gaps in areas
central to professional practice. SAKE, its evaluation scripts, and all results are released
as open source to give the community a baseline for tracking architectural reasoning in LLMs.

\keywords{Software Architecture  \and Software Architectural Knowledge \and Benchmark \and Large Language Model.}
\end{abstract}
\section{Introduction}
\label{sec:introduction}

The role of LLMs in software engineering has expanded rapidly, with practitioners now deploying them across tasks ranging from code completion to documentation and system design.
Yet as adoption grows, so does the need to understand where these models are genuinely competent and where they are not.
Software architecture represents one of the least examined areas in this regard: architectural decision-making demands a deep understanding of quality attribute trade-offs, design patterns, and system-level constraints, reasoning that goes well beyond the syntactic and algorithmic problem-solving that most existing benchmarks measure~\cite{chen2021humaneval,hendrycks2021mmlu,srivastava2022bigbench}.
As a result, practitioners currently lack a principled basis for deciding when to trust an LLM as an architectural advisor, and researchers have no shared baseline against which to measure progress.

A systematic review by Schmid et al. identified four main application areas of LLMs in software architecture, yet found that approximately one-third of studies do not compare results against any baseline~\cite{schmid25}.
Soliman and Keim explicitly acknowledge that LLMs have not been previously evaluated for their ability to answer questions about architectural knowledge, leaving doubts about their accuracy, quality, and trustworthiness~\cite{soliman25}.
To the best of our knowledge, no dedicated benchmark for this domain currently exists.

To address this gap, we introduce \textsc{SAKE} (\textit{Software Architectural Knowledge Evaluation}), the first standardized and reproducible benchmark explicitly designed to assess software architectural knowledge in LLMs.
\textsc{SAKE} comprises 2154 expert-curated multiple-choice questions stratified across eight core architectural categories, derived from two canonical references: \emph{Software Architecture in Practice}~\cite{bass21} and the Gang of Four Design Patterns~\cite{gamma95}.
Each question of the benchmark was independently authored and subjected to a dual peer review before inclusion, ensuring content validity and inter-rater agreement.
Questions are further stratified by context length, capturing whether model performance degrades as prompts require broader contextual integration.

We use \textsc{SAKE} to benchmark 11 state-of-the-art LLMs, covering both proprietary and open-weight models, in zero-shot and five-shot settings.
Our evaluation reveals that overall accuracy is consistently high, but performance varies substantially across architectural knowledge categories, with Architectural Solutions consistently among the most challenging. We further find that the effect of prompt context length is category-dependent: it tends to help on recall-oriented categories while degrading accuracy on reasoning-heavy ones such as Architectural Solutions.

SAKE assesses whether a model holds and can apply established architectural knowledge, which is a prerequisite for sound design support but not a substitute for it. It does not measure the quality of open-ended design decisions on a concrete system, where context, competing drivers, and trade-off judgment dominate. We therefore position SAKE as a foundational knowledge layer: a model that fails here cannot be trusted to reason about architecture, while strong performance here is a necessary first filter rather than a guarantee of good architecting.

This work is guided by three research questions:

\begin{description}
    \item[\textbf{RQ1}] \textit{How do LLMs perform on software architectural
    knowledge overall, in both zero-shot and five-shot settings?}

    \item[\textbf{RQ2}] \textit{How does LLM performance vary across 
    software architectural knowledge categories, in both zero-shot and
    five-shot settings?}

    \item[\textbf{RQ3}] \textit{How does prompt context length affect LLM
    performance on software architectural knowledge?}
\end{description}

The paper is organized as follows: 
Section~\ref{sec:background} introduces the background and related works.
Section~\ref{sec:meth} details the methodology used in this work.
Section~\ref{sec:sake} details the construction of the benchmark.
Section~\ref{sec:setup} and Section~\ref{sec:results} show the settings, how the benchmarks were configured, the obtained results, the discussion, and the implications for researchers and practitioners.
Finally, we report the current limitations of SAKE in Section~\ref{sec:limitation} and draw the conclusions in Section~\ref{sec:conclusion}.

The \textbf{supplemental materials} containing the code and the benchmark are available at the following link~\footnote{\url{https://anonymous.4open.science/r/SAKE-1DD7/README.md}}.

\section{Background and Related work}~\label{sec:background}
\subsection{Software Architectural Knowledge}

Software architecture defines the fundamental structural organization of a
system through its components, connectors, and the constraints governing their
interactions~\cite{garlan1994,bass21,gamma95,perry92}.
Unlike lower-level design concerns, architectural decisions are made early in
the development lifecycle and carry consequences that are costly and difficult to reverse~\cite{babar2005}.
The knowledge required to make sound architectural choices is broad and complex, spanning architectural patterns, quality attribute trade-offs, design tactics, and decision rationale~\cite{farshidi2020}.

Perry and Wolf established that architecture encompasses elements, form, and
rationale, positioning the reasoning behind design choices as equally important
as the choices themselves~\cite{perry92}.
This gave rise to Architectural Knowledge Management, which views architects as
decision-makers balancing competing quality concerns, a capability that demands
both declarative knowledge from literature and episodic knowledge from
practice~\cite{capilla2016,babar2005}.

A persistent challenge is that a substantial portion of Software architectural Knowledge (SAK) remains tacit and
undocumented: rationale regarding middleware selection, API choices, or quality
attribute prioritization is typically lost when architects leave a
project~\cite{avgeriou2007}.
The explicit knowledge that does exist is equally fragmented; a systematic
review by Farshidi et al.\ found that knowledge on patterns and their quality
attribute impacts was scattered across more than 200 publications with no
consolidated basis for decision-making~\cite{farshidi2020}.
This breadth, tacitness, and fragmentation collectively explain why architectural
reasoning constitutes a distinctly demanding cognitive task that cannot be
reduced to syntactic or algorithmic problem-solving.

\subsection{Large Language Models in Software Architecture}

LLMs are increasingly applied to software architecture tasks.
A systematic literature review by Schmid et al.\ analyzed 18 relevant articles
and identified four application categories: reference architectures,
classification and detection, extraction and generation, and assistants.
It found that 73\% of studies rely on decoder-only GPT-based models with
zero-shot prompting as the dominant strategy in 70\% of cases, while advanced
techniques such as Retrieval-Augmented Generation and Chain-of-Thought prompting
remain largely underexplored~\cite{schmid25}.

On architectural knowledge specifically, Soliman and Keim queried GPT-3.5 about
the Hadoop HDFS system with 14 software engineers, obtaining a moderate average
recall of 0.594 but a lower precision of 0.395, with engineers rating overall
quality and trustworthiness as only moderate~\cite{soliman25}.
Cervantes et al.\ showed that LLMs can produce architectures closely aligned
with established ADD-based solutions, though continuous human oversight remains
necessary to address hallucinations and mixed abstraction levels~\cite{cervantes25}.
Amalfitano et al.\ evaluated four LLMs on architecture recovery from source
code, finding strong performance on broad structural elements and architectural
styles, but notable struggles with fine-grained design patterns and complex
inter-class relationships~\cite{amalfitano25}.

\subsection{Benchmarks for Large Language Models}

Standardized benchmarks have become essential tools for measuring and comparing
LLM capabilities across diverse tasks.
MMLU~\cite{hendrycks2021mmlu} covers 15,908 multiple-choice questions across
57 subjects, estimating human expert-level accuracy at 89.8\%.
BIG-Bench~\cite{srivastava2022bigbench} extends this with 204 collaborative
tasks across mathematics, common-sense reasoning, social bias, and software
development.
HumanEval~\cite{chen2021humaneval} targets code generation specifically, with
164 hand-written programming problems evaluated via the pass@k metric, i.e., the probability that at least one out of k generated code samples passes all unit tests of the problem. HELM~\cite{liang2023helm} takes a broader view, evaluating models across 42
scenarios using seven metrics including accuracy, fairness, robustness, and
toxicity.
However, a 2026 systematic study of 60 benchmarks found that nearly half
exhibit high saturation and can no longer effectively differentiate between
top-performing models, with expert-curated benchmarks showing greater longevity than crowdsourced ones~\cite{saturation2026}.

\paragraph{Gap Analysis.} Despite this broad landscape of evaluation frameworks, a critical gap remains in the 
domain of software architecture. While LLMs are increasingly applied to architectural tasks such as design decision classification, architecture recovery, and knowledge extraction~\cite{schmid25}, no dedicated benchmark exists for evaluating their performance on software architectural knowledge.
Soliman and Keim explicitly acknowledge that their exploratory case study on GPT and HDFS represents only a first baseline, noting that LLMs have not been previously evaluated for their ability to answer questions about architectural knowledge, leaving doubts about their accuracy, quality, and 
trustworthiness~\cite{soliman25}. 
Furthermore, Schmid et al.\ report that approximately one-third of studies in this area do not compare results against any baseline, and that entire areas such as conformance checking and the evaluation of quality attributes like evolvability remain completely unaddressed~\cite{schmid25}. 
This work aims to contribute to filling this gap by proposing a structured benchmark specifically designed to assess LLM capabilities across the core concepts of Software Architectural Knowledge.

\section{Methodology}
\label{sec:meth}
\begin{figure}[h]
    \centering
    \includegraphics[width=\columnwidth]{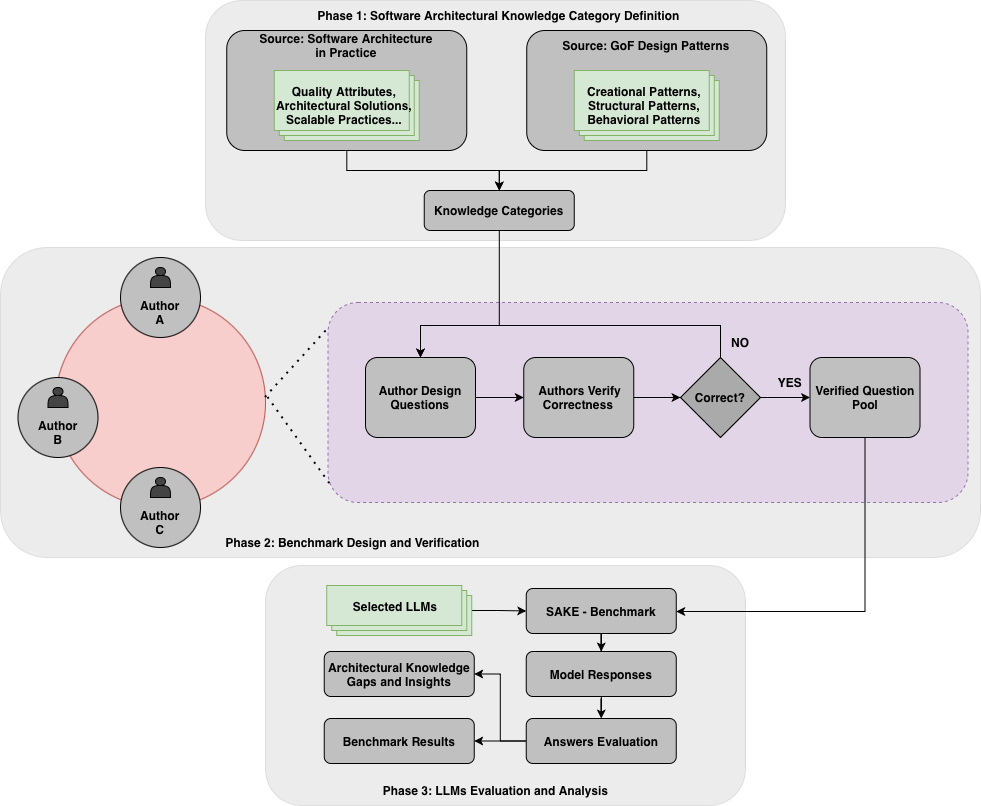}
    \caption{SAKE Benchmark Methodology}
    \label{fig:Meth1}
\end{figure}

This work's methodology is organized into three sequential phases, as shown in Figure~\ref{fig:Meth1}.
The \textbf{first phase} focuses on defining the software architectural knowledge categories that structure the entire benchmark. We draw the software architecture categories from two canonical reference works: \textit{Software Architecture in Practice}~\cite{bass21}, which covers topics like quality attributes, architectural solutions, and scalable practices, and the Gang of Four \textit{Design Patterns}~\cite{gamma95}, which 
contributes creational, structural, and behavioral patterns.

The resulting set of knowledge categories forms the backbone of SAKE and is described in detail in Section~\ref{sec:categories}.

We selected these two works for three reasons. First, both are canonical and widely
taught, which gives the benchmark a transparent and defensible scope. Second, together
they span two complementary levels of architectural reasoning: system-level concerns
such as quality attributes, tactics, and styles from Bass et al., and the reusable
object-level design vocabulary from the Gang of Four that architects still rely on when
reasoning about component-level structure. Third, both have stable, well-defined chapter
structures that map cleanly onto knowledge categories, which supports balanced coverage.

We stress that the two works\cite{bass21,gamma95} are used at this stage only, to define the category taxonomy.
They are not a question bank: no item is copied or paraphrased from their text. The
questions themselves are produced in the second phase by the authors.

The \textbf{second phase} covers benchmark design and verification. Each of the three authors independently designs questions grounded in the defined categories, which are then subject to a dual independent review: the other two authors assess every question for correctness and quality, following a content validation approach aimed at ensuring inter-rater agreement before inclusion in the final pool. 
Questions that do not pass this review step are revised and resubmitted, while those that receive approval from both reviewers are added to the verified question pool. 
The full question construction process is presented in Section~\ref{sec:construction}.

The \textbf{third phase} uses the verified benchmark to evaluate a set of selected LLMs. Each model is prompted with the benchmark questions, and its responses are collected and scored. The evaluation produces two main outputs: benchmark performance results and a set of architectural knowledge 
gaps and insights that highlight where current LLMs fall short. 
The experimental setup is described in Section~\ref{sec:setup}, and the results are reported and discussed in Section~\ref{sec:results}.

\section{The SAKE Benchmark}
\label{sec:sake}
\subsection{Design Goals and Principles}

SAKE was designed from the ground up to address the specific gaps identified in the related work. Five core principles guided every design decision, from the choice of knowledge categories to the question construction process.

\noindent\textbf{i) Domain specificity.} Existing benchmarks such as MMLU~\cite{hendrycks2021mmlu} and HumanEval~\cite{chen2021humaneval} measure general language understanding and code generation, but neither captures the reasoning demands specific to software architecture. 
Architectural decision-making requires integrating knowledge of quality attribute trade-offs, design patterns, and system-level constraints in ways that syntactic or algorithmic tasks simply do not exercise~\cite{bass21,gamma95}. SAKE is therefore scoped exclusively to Software Architectural Knowledge, ensuring that evaluation results reflect software architectural competency.

\noindent\textbf{ii) Grounding in established knowledge sources.} To define what architectural knowledge a benchmark should cover, we anchor the category space in two canonical references widely recognized by both academia and industry: \textit{Software Architecture in Practice}~\cite{bass21}, and the Gang of Four \textit{Design Patterns}~\cite{gamma95}. Founding the benchmark in these sources provides a transparent and defensible scope, making clear to the community what SAKE measures and what it does not.

\noindent\textbf{iii) Expert curation and content validity.} One concern identified in the benchmark saturation literature is that crowdsourced question sets tend to degrade in discriminative power over time, while expert-curated benchmarks retain longevity~\cite{saturation2026}. In SAKE, each question is designed by a domain expert and independently reviewed by the other authors before inclusion. This dual-review process enforces inter-rater agreement and filters out ambiguous or misleading items, ensuring that every question reflects a test of architectural understanding rather than a test of reading comprehension or phrasing interpretation.

\noindent\textbf{iv) Standardization and reproducibility.} A benchmark is only useful if it can be run under identical conditions, thus making results comparable across studies. SAKE adopts a uniform multiple-choice format with exactly four options per question, which eliminates scoring ambiguity and supports automated evaluation at scale. All questions, evaluation scripts, and results are released as open-source artifacts, so any team can replicate or extend our experiments without re-implementing the evaluation pipeline.

\noindent\textbf{v) Granularity.} Rather than reporting a single aggregate score, SAKE is stratified across eight architectural knowledge categories and four context-length levels. This structure makes it possible to identify not just how well a model performs overall, but where its architectural competency is strong and where it falls short. Diagnostic resolution of this kind allows the benchmark to generate actionable insights for both practitioners deciding whether to trust an LLM as an architectural advisor and researchers working to improve model capabilities in targeted areas.

\subsection{Software Architectural Knowledge Categories}
\label{sec:categories}

SAKE is divided into eight knowledge categories, derived from two canonical references that together cover the breadth of software architectural knowledge. Three categories come from the chapters of Gang of Four \textit{Design Patterns}~\cite{gamma95}, which provides the foundational vocabulary for reusable object-oriented design. The remaining five categories come from the chapters of \textit{Software Architecture in Practice}~\cite{bass21}, which addresses system-level concerns such as quality attributes, architectural styles, and emerging computing paradigms. 

We include the three design-pattern categories deliberately: although they operate below
the level of architectural styles, pattern knowledge is part of the working vocabulary an
architect uses when reasoning about component structure, and keeping them as separate
categories lets us measure that knowledge in isolation rather than conflating it with
system-level reasoning.

Table~\ref{tab:categories} summarizes the eight categories, their source, and their frequency in the benchmark.

\begin{table}[h]
\scriptsize
\centering
\caption{SAKE knowledge categories grouped by source.}
\label{tab:categories}
\begin{tabular}{l|l|c}
\hline
\textbf{Category Source} & \textbf{Knowledge Category} & \textbf{\#Questions} \\
\hline
Design Patterns~\cite{gamma95} & Creational Patterns (CP) & 319 \\
Design Patterns~\cite{gamma95} & Structural Patterns (SPA)  & 222 \\
Design Patterns~\cite{gamma95} & Behavioral Patterns (BP)  & 280 \\
Software Architecture in Practice~\cite{bass21} & Quality Attributes (QA) & 290 \\
Software Architecture in Practice~\cite{bass21} & Architectural Solutions (ASO) & 251 \\
Software Architecture in Practice~\cite{bass21} & Scalable Practices (SPR) & 264 \\
Software Architecture in Practice~\cite{bass21} & Architectural Styles (AST) &  263\\
Software Architecture in Practice~\cite{bass21} & SA for Quantum Computing (QC) & 265 \\
\hline
\end{tabular}
\end{table}

\noindent\textbf{Creational Patterns (CP)} cover object creation mechanisms that decouple a system from the specifics of how its objects are instantiated and composed. Questions in this category assess knowledge and reasoning on patterns such as Factory Method, Abstract Factory, Builder, Prototype, and Singleton, testing whether a model understands not only the mechanics of each pattern but also the conditions under which one is preferred over another.

\noindent\textbf{Structural Patterns (SPA)} concern how classes and objects are composed to form larger, more flexible structures. This category includes patterns such as Adapter, Bridge, Composite, Decorator, Facade, Flyweight, and Proxy, each of which addresses a recurring challenge in managing dependencies and simplifying interactions between system components.

\noindent\textbf{Behavioral Patterns (BP)} address the assignment of responsibilities between objects and the communication protocols between them. Patterns such as Observer, Strategy, Command, Iterator, and Template Method belong to this category, which tests whether a model can reason about dynamic behavior and control flow at the design level.

\noindent\textbf{Quality Attributes (QA)} test the model's ability to reason about non-functional requirements such as performance, availability, security, modifiability, and testability. Understanding quality attributes is central to architectural decision-making, since these properties govern the trade-offs that architects must balance throughout the system lifecycle.

\noindent\textbf{Architectural Solutions (ASO)} cover the tactics and design decisions that architects employ to satisfy quality attribute requirements. Rather than testing abstract definitions, questions in this category ask models to connect architectural choices to their expected effects on system properties, reflecting the kind of reasoning that practitioners apply when selecting between competing solutions.

\noindent\textbf{Scalable Practices (SPR)} address design strategies and organizational patterns supporting the growth of a system in terms of load, data volume, team size, or deployment complexity. This category evaluates whether a model understands architectural approaches commonly adopted in large-scale distributed systems, including decomposition strategies, stateless design, and data partitioning.

\noindent\textbf{Architectural Styles (AST)} cover the structural paradigms defining how a system's components are organized and interact, including layered, event-driven, microservices, pipe-and-filter, and service-oriented architectures. Evaluating this category reveals whether a model correctly characterizes each style and reasons about its suitability for different system requirements and constraints.

\noindent\textbf{Software Architecture and Quantum Computing (QC)} is the most forward-looking category in SAKE, reflecting the growing relevance of quantum computing to system design. Questions here probe knowledge of architectural considerations unique to quantum systems, such as hybrid classical-quantum integration, qubit management, and the constraints imposed by near-term quantum hardware on system design decisions.

We include this category deliberately as a forward-looking probe. Hybrid classical-quantum
systems are beginning to appear in the architecture literature, and a knowledge benchmark
is the right place to establish an early baseline before such systems become common.

\subsection{Question Construction Methodology}
\label{sec:construction}

Our procedure follows the logic of expert content validation used in established
multiple-choice benchmarks~\cite{hendrycks2021mmlu,medqa}: independent authoring, dual
expert review against a fixed rubric (correctness, clarity, category fit), and consensus
resolution of disagreements before an item enters the pool.

Building a benchmark that tests architectural reasoning requires careful attention to both how questions are written and how they are validated.
The construction of the SAKE question pool followed a structured authoring and review process that prioritized content validity and balanced coverage across categories.

Each of the three authors independently designed questions grounded in the eight knowledge categories defined in Section~\ref{sec:categories}. The authors used the two source books to fix the eight categories, and then authored
original questions within each category from their own domain expertise. Items are not
extracted from the books; instead, dual review checks that each question and its key are correct with respect to established architectural knowledge, which keeps content grounded
while leaving the wording original. Questions were distributed as evenly as possible across the eight categories, so that no single area dominates the benchmark and all facets of Software Architectural Knowledge receive comparable coverage.

Every question in SAKE follows a uniform multiple-choice format with exactly four answer options, consistent with widely adopted benchmarks~\cite{hendrycks2021mmlu,medqa}.
Each question has one unambiguously correct answer, while the remaining three options are designed to be plausible rather than trivially wrong. Plausible distractors are important because they require a model to reason carefully about the distinctions between related concepts, rather than eliminating incorrect options through superficial pattern matching. This design choice directly supports the goal of measuring architectural understanding.

SAKE questions are also stratified by context length, defined as the total number of words in the question.
After the dataset design step, we partitioned the questions into four bins based on the quartiles of the word count distribution of the benchmark: short (0--134 words), medium-short (135--195 words), medium-long (196--265 words),
and long (more than 266 words), corresponding to the 25th, 50th, and 75th percentile boundaries, respectively. 
We decided to count the number of words instead of the number of tokens because different models use different tokenization algorithms.
Each question is optionally enhanced via an LLM; specifically, for longer questions, the context is augmented to increase the complexity of the author's original scenario.

Once a question was authored, it was submitted to a dual independent review by the other authors, who assessed it for correctness, clarity, and category fit. This review process follows a 
content validation approach aimed at achieving inter-rater agreement before any question is admitted to the final pool. If both reviewers approved the question, it was added directly to the verified pool. If one or both reviewers raised concerns, the question was not discarded but returned to the author for revision. Disagreements were resolved through discussion among all three authors until a consensus was reached, ensuring that every admitted question reflects a shared judgment of quality rather than a single author's perspective.

Of the 2191 questions initially authored, 2094  were accepted by both reviewers without
change, 60 were returned at least once for revision, and 37 were discarded as
ambiguous or out of scope, reaching a first-pass agreement of 95.6\%. 

The process yielded a final pool of 2154 verified questions. The iterative nature of the review cycle, combined with the breadth of eight architectural categories and four context-length levels, 
required several months to complete. 
The resulting benchmark provides a principled, expert-validated basis for evaluating LLM performance on Software Architectural Knowledge.

Representative items for each category are provided in Appendix~\ref{app:examples}.

\section{Experimental Setup}~\label{sec:setup}
We evaluate 11 LLMs on SAKE using zero-shot and five-shot prompting.
All experiments are conducted using the OpenRouter API in Python 3.14, with the temperature fixed to 0 to ensure more deterministic outputs.

We benchmark a set of widely used models across multiple providers, selected based on their popularity and ability to generate concise outputs compatible with multiple-choice answering.
The evaluated models include  Claude Haiku 4.5, Claude Sonnet 4.6, Claude Opus 4.6, Deepseek v3.2, Gemini 3.1 Flash Lite, GPT 5.4 Nano, GPT 5.4, Grok 4.1 Fast, Grok 4.2, Mistral Small 4, and Qwen3 235B A22B.

\paragraph{\textbf{Prompting and Metrics}}
All prompts follow the format:
\begin{small}
\begin{verbatim}
{question}
A. {choice_0}
B. {choice_1}
C. {choice_2}
D. {choice_3}

Answer with a single letter (A, B, C, or D).
Answer:
\end{verbatim}
\end{small}
In the five-shot setting, five examples belonging to the same category are prepended to the query, using identical formatting to the zero-shot prompt and including the correct answer. 
These examples have been randomly selected from the dataset and removed from the evaluation set, resulting in a total of 40 held-out queries (5 per category), stored in the `validation' partition of the benchmark to ensure reproducibility.
Of the 2154 verified items, 40 (5 per category) are held out as five-shot examples, leaving
2114 scored test items. Zero-shot and five-shot are scored on the same 2114 items so that the
two settings are directly comparable.
To ensure consistency across models and simplify answer extraction, we constrain the outputs to 10 tokens, with the exception of GPT models, where the limit is 16 tokens due to API constraints.
This restriction forces concise responses, reducing the variability in output formatting, thus improving comparability across models.
Predicted answers are extracted via a cascade of regular expressions, following a procedure similar to~\cite{mmlupro}.
We first match the expected format specified in the prompt (`\verb|X|', where X is the answer's letter). If no match is found, we apply a secondary pattern that handles different formatting variations (e.g.,`\verb|**X**|').
As a final fallback, we extract the last standalone option letter among \{A, B, C, D\}. If all extraction steps fail, we assign a random label.

We define accuracy as the number of items for which the extracted answer matches the verified
key, divided by the number of items evaluated:
$\mathrm{Acc} = \frac{1}{N}\sum_{i=1}^{N}
\mathbf{1}[\hat{y}_i = y_i]$. 
We report it overall, per category, and as the unweighted mean
across categories.
Both the prompt design and the regex-based answer extraction mechanism are made open-source.


\section{Results}
\label{sec:results}
  \subsection{RQ1: Overall LLM Performance on SAKE}

\begin{table}[ht]~\caption{Overall accuracy on SAKE.}~\label{tab:overall_accuracy}
\centering
\scriptsize
\begin{tabular}{l|c|c}
\textbf{Model}        & \textbf{Zero-shot Acc{[}\%{]}} & \textbf{Five-shot Acc{[}\%{]}} \\ \hline
Claude Haiku 4.5      & 92.15                          & 93.71                          \\
Claude Sonnet 4.6     & 93.71                          & 93.8                           \\
Claude Opus 4.6       & 93.57                          & \textbf{94.23}                 \\
Deepseek v3.2         & 93.61                          & 89.4                           \\
Gemini 3.1 Flash Lite & 93.66                          & 93.14                          \\
GPT 5.4 Nano          & 90.96                          & 89.31                          \\
GPT 5.4               & \textbf{93.95}                 & 94.13                          \\
Grok 4.1 Fast              & 92.05                          & 92.57                          \\
Grok 4.2              & 93.28                          & 93.14                          \\
Mistral Small 4       & 92.48                          & 91.06                          \\
Qwen 3 235B A22B      & 93.38                          & 93.0                          
\end{tabular}
\end{table}

Table~\ref{tab:overall_accuracy} reports the overall accuracy obtained by each model on SAKE, both in the zero-shot and five-shot setting.

On the zero-shot scenario, performance is consistently high across all evaluated models, with the lowest accuracy being 90.96\%, achieved by GPT 5.4 Nano.
The second lowest accuracy is obtained by Grok 4.1 Fast, scoring 92.05\%. 
The top three performing models are GPT 5.4, Claude Sonnet 4.6, and Gemini 3.1 Flash Lite, yielding 93.95\%, 93.71\%, and 93.66\%, respectively.
Notably, Claude Sonnet 4.6 achieves slightly better performance than Claude Opus 4.6, outperforming it by 0.14\% in accuracy, even at a lower price ($3$ USD vs $5$ USD per million input tokens).
In the five-shot scenario, the increased prompt length leads to variable effects across models.
Several models exhibit degradation in performance when compared to their zero-shot counterparts.
Specifically, Deepseek v3.2 shows the largest drop, decreasing from 93.61\% to 89.4\% accuracy (-4.21\%). 
Smaller but consistent decreases are also shown by Gemini 3.1 Flash Lite (-0.52\%), GPT 5.4 Nano (-1.65\%), Mistral Small 4 (-1.42\%), Grok 4.2 (-0.14\%), and Qwen3 235B A22B (-0.38\%).
In contrast, a subset of the models benefits from the additional context.
Claude Opus 4.6 achieves the highest overall accuracy on SAKE at 94.23\%, improving by 0.66\% over its zero-shot performance.
Similar but slightly more contained improvements are observed for GPT 5.4 (94.13\%), Grok 4.1 Fast (92.57\%), Claude Sonnet 4.6 (93.80\%), and Claude Haiku 4.5 (93.71\%).
These results suggest that the effectiveness of the few-shot prompting in SAKE is strongly dependent on the model's ability to utilize longer input contexts. 
Higher capacity models are, in general, able to effectively use the in-context examples, while smaller ones rapidly degrade in performance.
Overall, the performance across both scenarios is high, ranging from 89.31\% (GPT 5.4 Nano, five-shot) to 94.23\% (Claude Opus 4.6, five-shot).
From a deployment perspective, Qwen 3 235B provides the most favorable accuracy-to-cost tradeoff. 
While it achieves competitive accuracy (within 1\% of the top performing models), its input token cost ($0.071\$$ USD per million tokens) is almost two orders of magnitude lower than Claude Opus 4.6 ($5$ USD per million tokens).
Additionally, its open-source availability makes it particularly suitable for on-premise or privacy-sensitive deployments.

\noindent\textbf{Answer to RQ1.} Overall accuracy is high and tightly clustered
(89.31--94.23\%), so on knowledge recall the strongest proprietary and open-weight models are
close to interchangeable. The practical differentiator is cost: Qwen~3~235B stays within one
point of the best models at roughly two orders of magnitude lower price, which makes raw
accuracy a weak basis for model choice on this task.

  \subsection{RQ2: Performance Across Software Architectural Knowledge Categories}

\begin{table}[ht]~\caption{Zero-shot per category accuracy [\%].}~\label{tab:zero_shot_per_cat}~\vspace{-4mm}
\centering
\scriptsize
\begin{tabular}{l|l|l|l|l|l|l|l|l|r}
\textbf{Model}        & \textbf{CP}   & \textbf{SPA}  & \textbf{BP}   & \textbf{QA}   & \textbf{ASO}  & \textbf{SPR}  & \textbf{QC}   & \textbf{AST}  & \textbf{Mean} \\ \hline
Claude Haiku 4.5      & 95.5          & 91.7          & 91.6          & 97.2          & 87.8          & 89.6          & 91.2          & 91.1          & 92.0          \\
Claude Sonnet 4.6     & \textbf{99.0} & 92.2          & 94.9          & \textbf{99.0}          & 89.8          & \textbf{93.0} & 89.6          & 89.9          & 93.4          \\
Claude Opus 4.6       & 98.4          & 91.7          & \textbf{95.3} & \textbf{99.0} & 90.2 & 92.7          & 89.2          & 89.9          & 93.3          \\
Deepseek v3.2         & 97.8          & 92.2          & 93.1          & \textbf{99.0}          & 90.2          & 91.1          & 92.3          & 91.5          & 93.4          \\
Gemini 3.1 Flash Lite & 97.1          & \textbf{92.6} & 94.6          & 97.9          & \textbf{91.1}          & 91.9          & 91.2          & 91.5          & 93.5          \\
GPT 5.4 Nano          & 95.5          & 90.8          & 91.3          & 97.2          & 86.6          & 88.4          & 85.0          & 91.1          & 90.7          \\
GPT 5.4               & 98.7          & 93.1          & 94.9          & 98.2          & 88.6          & 91.1          & \textbf{92.7} & \textbf{92.2} & \textbf{93.7} \\
Grok 4.1 Fast              & 95.2          & 89.4          & 92.7          & 97.5          & 90.2          & 88.4          & 90.0          & 91.1          & 91.8          \\
Grok 4.2              & 97.1          & 91.7          & 93.1          & 97.9          & 90.2          & 92.7          & 91.5          & 90.3          & 93.1          \\
Mistral Small 4       & 96.2          & 91.7          & 92.4          & 98.2          & 89.8          & 89.6          & 89.2          & 91.1          & 92.3          \\
Qwen 3 235B A22B      & 97.8          & \textbf{92.6}          & 94.9          & 98.6          & 89.8          & 90.4          & 91.5          & 89.5          & 93.2          \\ \hline
Mean                  & 97.1 & 91.8          & 93.5          & 98.2          & 89.5          & 90.8          & 90.3          & 90.8          &              -
\end{tabular}
\end{table}

Table~\ref{tab:zero_shot_per_cat} reports the per-category accuracy and overall mean performance ($Mean$) in the zero-shot setting.
Across all models, \textit{QA} is the easiest category, with an average accuracy of 98.2\% and limited variance.
Even the weakest model, GPT 5.4 Nano, achieves 97.2\%, showing that modern models possess in-depth knowledge of Quality Attributes.
\textit{CP} follows with a high average of 97.1\%, with top models such as Claude Sonnet 4.6 reaching near perfect performance (99\%), with a modest gap with the lowest performer at 95.2\%.

On the other hand, \textit{ASO} (89.5\%) and \textit{QC} (90.3\%) are the most complex categories in our evaluation. These exhibit both lower average accuracy and higher deviation between models.
Notably, GPT 5.4 Nano obtains the lowest scores in both categories (86.6\% and 85\%), showing how the model scale impacts harder tasks.
The remaining categories (SPA, BP, SPR, AST) show intermediate behavior, with accuracies in the range 88\%-94\%, and moderate inter-model variability.
In terms of overall performance, GPT 5.4 achieves the highest mean accuracy (93.7\%), followed closely by Gemini 3.1 Flash Lite and Deepseek v3.2 (93.5\% and 93.4\% respectively).
Given their significantly lower cost per token ($2$ USD vs $0.26$ USD per million input tokens), these models represent competitive alternatives with little accuracy degradation ($<0.3\%$).

\begin{table}[ht]~\caption{Five-shot per-category accuracy [\%].}~\label{tab:five_shot_per_cat}
\scriptsize
\centering
\begin{tabular}{l|l|l|l|l|l|l|l|l|r}
\textbf{Model}        & \textbf{CP}   & \textbf{SPA}  & \textbf{BP}   & \textbf{QA}   & \textbf{ASO}  & \textbf{SPR}  & \textbf{QC}   & \textbf{AST}  & \textbf{Mean} \\ \hline
Claude Haiku 4.5      & 97.4          & 91.2          & 92.7          & 98.2          & 88.2          & 93.4          & \textbf{95.0} & \textbf{91.5}          & 93.5          \\
Claude Sonnet 4.6     & 99.0          & 92.6          & \textbf{95.3}          & 98.6          & 89.8          & 93.0          & 89.6          & 90.3          & 93.5          \\
Claude Opus 4.6       & 98.7          & 91.7          & \textbf{95.3} & \textbf{99.0} & \textbf{90.6} & \textbf{95.0} & 91.2          & 90.3          & \textbf{94.0} \\
Deepseek v3.2         & 98.4          & 78.3          & 93.1          & 98.6          & 85.4          & 84.2          & 88.1          & 84.1          & 88.8          \\
Gemini 3.1 Flash Lite & 97.8          & 91.2          & 93.1          & 98.2          & 89.8          & 92.3          & 90.8          & 89.9          & 92.9          \\
GPT 5.4 Nano          & 89.8          & 88.9          & 90.9          & 95.1          & 83.7          & 89.6          & 87.3          & 88.0          & 89.2          \\
GPT 5.4               & \textbf{99.4} & \textbf{93.6} & 94.6          & 98.2          & 89.8          & 90.7          & 93.8          & 91.1          & 93.9          \\
Grok 4.1 Fast             & 98.4          & 90.8          & 93.4          & 96.8          & 87.8          & 90.0          & 89.6          & \textbf{91.5} & 92.3          \\
Grok 4.2              & 97.4          & 92.2          & 94.2          & 96.8          & 89.4          & 93.0          & 91.2          & 89.2          & 92.9          \\
Mistral Small 4       & 89.5          & 92.2          & 93.1          & 97.9          & 89.0          & 86.9          & 90.0          & 89.5          & 91.0          \\
Qwen 3 235B A22B      & 97.8          & 92.2          & 94.2          & 98.2          & 88.6          & 89.6          & 91.5          & 89.9          & 92.8          \\ \hline
Mean                  & 96.7          & 90.4          & 93.6          & 97.8 & 88.4          & 90.7          & 90.7          & 89.6          & -            
\end{tabular}
\end{table}

Table~\ref{tab:five_shot_per_cat} reports the per-category accuracy in the five-shot setting.
Consistent with the zero-shot scenario, \textit{QA} and \textit{CP} remain the easiest categories, with average accuracies of 97.8\% and 96.7\% respectively.
In particular, peak accuracy is obtained by Claude Opus 4.6 and Claude Sonnet 4.6 on QA (99.0\% and 98.6\%) and by GPT 5.4 on CP (99.4\%).
In contrast, \textit{ASO} (88.4\%) and \textit{AST} (89.6\%) emerge as the most challenging categories. Compared to zero-shot, these tasks show increased instability, with multiple models significantly degrading in accuracy.
In particular, Deepseek v3.2 and GPT 5.4 Nano drop to 85.4\% and 83.7\% on ASO, and Deepseek further degrades on AST (84.1\%).
When comparing to zero-shot results, the main difference is the increased variance across categories and models. Top-tier models (Claude Sonnet/ Opus, GPT 5.4) remain stable, mid-range models designed for efficiency show inconsistent behavior.
In particular, Deepseek v3.2 experiences a sharp drop in SPA (78.3\%) and SPR (84.2\%), showing its sensitivity to longer prompts.
Overall, Claude Opus 4.6 achieves the highest mean accuracy (94\%), closely followed by GPT 5.4 (93.9\%).
However, the improvement w.r.t. lower-cost alternatives (e.g., Gemini~3.1~Flash~Lite at 92.9\%) is limited, considering in particular the sharp increase in price per query of the five-shot scenario.

\noindent\textbf{Answer to RQ2.} Competence is uneven. Every model is near the ceiling on Quality Attributes and Creational Patterns, whereas Architectural Solutions is consistently the weakest category; Quantum Computing is also among the hardest in the zero-shot setting, while in five-shot the second-weakest category shifts to Architectural Styles. Adding five-shot examples widens rather than narrows the spread for efficiency-oriented models. A single aggregate score therefore hides where a model can and cannot be trusted, which is why we report and recommend consulting results per category.

\subsection{RQ3: Performance Across Context Length}
\begin{table}[ht]~\caption{Accuracy at different prompt lengths (\#words).}~\label{tab:acc_per_word}
\centering
\scriptsize
\begin{tabular}{l|l|l|l|l|l|l|l|l}
                      & \multicolumn{4}{l}{\textbf{Overall Acc{[}\%{]}}}                                  & \multicolumn{4}{l}{\textbf{ASO Acc{[}\%{]}}}                                      \\ \hline
\textbf{Model}        & \textbf{0-134} & \textbf{135-195} & \textbf{196-265} & \textbf{$>$266} & \textbf{0-134} & \textbf{135-195} & \textbf{196-265} & \textbf{$>$266} \\
Claude Haiku 4.5      & 89.0           & 93.1             & 94.3             & 92.2                       & 94.9           & 89.6             & 87.0             & 84.2                       \\
Claude Sonnet 4.6     & 91.2           & 93.7             & 95.4             & 94.5                       & 94.9           & 93.8             & 88.3             & 86.6                       \\
Claude Opus 4.6       & 90.4           & 93.5             & 95.3             & \textbf{95.1}              & 94.9           & 95.8    & 87.0             & \textbf{87.8}              \\
Deepseek v3.2         & 90.3           & 93.3             & \textbf{95.8}             & \textbf{95.1}                       & 97.4           & 93.8             & 87.0             & \textbf{87.8 }                      \\
Gemini 3.1 Flash Lite & 90.4           & \textbf{93.9}    & \textbf{95.8}    & 94.5                       & 97.4           & 95.8             & 88.3             & \textbf{87.8}                       \\
GPT 5.4 Nano          & 86.1           & 91.2             & 93.9             & 92.6                       & 89.7           & 91.7             & 84.4             & 84.2                       \\
GPT 5.4               & \textbf{92.0}  & 93.3             & \textbf{95.8}             & 94.7                       & 94.9           & 89.6             & 87.0             & 86.6                       \\
Grok 4.1 Fast             & 88.4           & 90.4             & 95.4             & 94.0                       & \textbf{100}  & 91.7             & 87.0             & \textbf{87.8}                       \\
Grok 4.2              & 89.5           & \textbf{93.9}             & 95.4             & 94.3                       & 92.3           & \textbf{97.9}             & \textbf{89.6}    & 85.4                       \\
Mistral Small 4      & 89.1           & 92.5             & 94.3             & 94.0                       & 94.9           & 93.8             & 88.3             & 86.6                       \\
Qwen 3 235B A22B      & 91.0           & 93.1             & 95.4             & 94.0                       & 97.4           & 93.8             & 87.0             & 86.6                      
\end{tabular}
\end{table}

Table~\ref{tab:acc_per_word} reports the model accuracy across different prompt length bins for the zero-shot setting.
Prompts following the format detailed in Section~\ref{sec:setup} are partitioned into four bins based on quartiles of their word count distribution.
For each bin, we report both the overall accuracy and accuracy on the ASO domain, as it represents the most complex category (lowest mean accuracy across models) for the zero-shot scenario (as reported in Table~\ref{tab:zero_shot_per_cat}).

On the overall dataset, shorter prompts (0-134 words) yield lower accuracy (Overall Acc) across all models, hinting that limited context information makes general software architecture questions more complex.
In this regime, GPT 5.4 yields the top performance (92\%), while GPT 5.4 Nano shows the lowest accuracy (86.1\%), highlighting the sensitivity of smaller models with limited context.
As the prompt length increases, performance improves across all models, with accuracies ranging from 90\% to 96\%. In the 135-195 word range, Gemini 3.1 Flash Lite achieves the top score (93.9\%), while in the 196-265 range multiple models, including GPT 5.4, Deepseek v3.2, and Gemini 3.1 Flash Lite, reach similar peak performance (95.8\%).
For the longest prompts ($>$266 words), Claude Opus 4.6 obtains the highest accuracy at 95.1\%, although with marginal improvements w.r.t. other models.
Notably, in the zero-shot setting, Deepseek v3.2 maintains competitive performance across all bins, without the degradation observed in the five-shot scenario, as the context length is still in the optimal range for this architecture.
A different trend emerges in the ASO category.
Shorter prompts lead to substantially higher accuracy, with Grok 4.1 Fast achieving perfect performance (100\%), followed by Qwen 3 235B, Deepseek v3.2, and Gemini 3.1 Flash Lite (all at 97.4\%). This hints that, for simpler or more direct queries within this domain, multiple models are able to rely on memorized or well-internalized knowledge.
In contrast, performance degrades as the prompt length increases. For instance, Grok 4.1 Fast drops from 100\% accuracy in the smallest bin to 91.7\% or less for prompts over 135 words.
A similar trend is shared across all models, indicating that longer and potentially more complex ASO queries introduce additional reasoning challenges. 
Within ASO, no single model dominates across length bins: Grok 4.1 Fast leads on the shortest prompts, Grok 4.2 attains the highest accuracy in both the 135–195 and 196–265 ranges (97.9\% and 89.6\%), and on the longest prompts (>266) four models, Claude Opus 4.6, Deepseek v3.2, Gemini 3.1 Flash Lite, and Grok 4.1 Fast, tie at 87.8\%.

The two trends run in opposite directions. On general knowledge questions, more context
raises accuracy (short prompts are the hardest), whereas on Architectural Solutions more
context lowers it. Quality Attributes, the easiest category, stays near ceiling at every
length, so the degradation is specific to reasoning-heavy categories. The likely cause is
distractor density rather than token budget: longer ASO stems introduce more competing but
plausible tactics, so the model must separate more near-correct options, which a larger
context window does not help with.

\noindent\textbf{Answer to RQ3.} Context length is not uniformly beneficial. It helps where
the task is recall and hurts where the task is trade-off reasoning. Adding context is a useful
default only outside the reasoning-heavy categories.

\subsection{Lessons learned.} First, high aggregate accuracy hides uneven competence,
so a single number is misleading for architectural use. Second, the strongest models are
cluster within a few points on knowledge, so selection should follow cost and category fit, not the
leaderboard. Third, the effect of context length is category-dependent: helpful for recall,
harmful for trade-off reasoning, which is the opposite of the "more context is always better"
intuition.
 \subsection{Implications for Practitioners}

Practitioners increasingly rely on LLMs as AI assistants for architectural tasks, yet our results show that this reliance carries meaningful risks that are not uniform across all areas of Software Architectural Knowledge. 
Models perform unevenly across the eight categories, meaning that a model that performs well in one area may fall short in another.
Practitioners should therefore treat LLM-generated architectural advice as category-dependent rather than uniformly reliable, and consult the SAKE category-level results to understand where a specific model is more or less trustworthy before deploying it in a given architectural context.

Our context-length analysis carries a practically important but non-uniform message: the effect of richer prompts depends on the type of architectural knowledge being tested. For most categories, longer prompts are associated with equal or higher accuracy so additional context generally aids recall-oriented tasks. The picture reverses in reasoning-heavy categories such as Architectural Solutions, where accuracy declines as prompts grow longer. This matters for practice because real architectural tasks rarely reduce to isolated facts; they require integrating information from system descriptions, requirement specifications, and competing quality-attribute scenarios. Practitioners should therefore not assume that more context uniformly improves an LLM's architectural judgment: while added detail tends to help on straightforward knowledge questions, it does not improve, and can even degrade, performance on complex, multi-constraint trade-off reasoning, which is where additional human scrutiny is most warranted.

\subsection{Implications for Researchers}

SAKE provides the research community with the first standardized and reproducible baseline for measuring LLM progress on Software Architectural Knowledge, establishing a common ground that was previously absent from the field. Researchers working to improve LLM architectural reasoning can use the category-level
results to identify the specific areas where current models fall short, allowing targeted improvements rather than undifferentiated pretraining or fine-tuning efforts.
The context-length performance gap points to a concrete research direction: improving the ability of models to integrate and reason over longer architectural scenarios. Techniques such as retrieval-augmented generation and chain-of-thought prompting remain largely underexplored in the 
architectural domain~\cite{schmid25}, and the structured nature of SAKE makes it well-suited to measure the impact of such approaches in a controlled and comparable way. Finally, since SAKE is fully open-source, researchers can extend it with new categories, additional questions, or 
updated model evaluations as the field evolves, ensuring that it serves as a living benchmark rather than a static snapshot.

\section{Threats to Validity}
\label{sec:limitation}

\noindent\textbf{Construct Validity.}

A multiple-choice format, while standardized and reproducible, cannot capture the full complexity of real architectural decision-making, which often involves open-ended trade-off analysis and context-specific judgment. To mitigate this, we grounded all questions in two canonical reference works and required each question to pass a dual independent review before inclusion, reducing the risk that questions test surface recall rather than genuine understanding. 
The reviewer pool was limited to the three authors. To reduce single-team bias, we
release the full question set openly and invite community review through the repository,
and we treat external validation as the natural next step for the benchmark.
An additional limitation concerns the benchmark's scope, which is constrained to the eight selected categories. Although these categories are broad, they may not capture the full breadth of software architectural knowledge. 

\noindent\textbf{Internal Validity.}

LLMs are sensitive to prompt formulation, and results may shift if questions are rephrased or presented in a different order. We address this by using a fixed, uniform prompt template across all models and runs.

\noindent\textbf{External Validity.}
 
While we selected a diverse set of both proprietary and open-weight models to maximize coverage, the landscape of LLMs evolves rapidly, and new models may exhibit different competency profiles. The findings should therefore be interpreted as a snapshot of the current state of LLM architectural reasoning rather than a permanent characterization of any model family. SAKE is fully open-source precisely to allow the community to re-run evaluations as new models emerge, ensuring that the benchmark remains a living baseline for tracking progress over time.

\section{Conclusions}
\label{sec:conclusion}
We introduce SAKE, a benchmark for software architectural knowledge comprising 2154 multiple-choice questions across eight categories, from Quality Attributes to more recent ones such as Quantum Computing.
The benchmark has been designed to assess the knowledge and reasoning capabilities of modern large language models in the software architecture domain.
By leveraging a structured multiple-choice format, SAKE enables automated, quantitative, and reproducible evaluation, reducing the reliance on manual, small-scale qualitative assessment common in modern works in the domain.

We evaluate 11 state-of-the-art models in zero-shot and five-shot settings. 
Results show consistently high overall accuracy, ranging from 89.31\% to 94.23\%, while also revealing non-uniform performance across domains.
In particular, we show how prompt design and longer context length affect the accuracy depending on the specific category of the question, as additional information in complex scenarios does not always translate to better accuracy.

Overall, SAKE represents an initial step toward a standardized and scalable evaluation framework for software architectural knowledge in large language models, and we expect it to gradually grow to feature more categories or different input data (e.g., diagrams) in the future. 

\bibliographystyle{splncs04}
\bibliography{biblio}

\appendix
\section{Example Questions}
\label{app:examples}
Table~\ref{tab:examples} gives one representative item per category. Each shows the
stem, the four options, and the verified key. The full set is available in the supplementary repository.

\scriptsize
\begin{longtable}{p{2.2cm}|p{8.8cm}}
\caption{One verified example per knowledge category. Correct option in bold, for readability in this subset c is always the correct answer.}
\label{tab:examples} \\
\hline
\textbf{Category} & \textbf{Example} \\
\hline
\endfirsthead

\multicolumn{2}{c}%
{{\tablename\ \thetable{} -- continued from previous page}} \\
\hline
\textbf{Category} & \textbf{Example} \\
\hline
\endhead

\hline
\multicolumn{2}{r}{{Continued on next page}} \\
\endfoot

\hline
\endlastfoot

Creational (CP) & In a distributed microservices environment where a client must instantiate a family of highly correlated configuration objects guaranteed to be mutually compatible based on a dynamic runtime topology graph, but the exact concrete classes are determined by an external service mesh discovery mechanism, which creational pattern best enforces the invariant that objects from different configuration families are never mixed during a single transactional scope? Options: (a) Builder with a Director, by utilizing a strict director class that enforces the construction sequence of the objects, ensuring compatibility through rigid construction pipelines. (b) Prototype Registry, by maintaining a centralized store of pre-configured instances that the client clones, relying on the registry's internal validation to prevent cross-family contamination. (c) \textbf{Abstract Factory, by encapsulating the instantiation of the entire family behind a single factory interface resolved for the specific topology, ensuring the client cannot accidentally mix concrete types.} (d) Parametrized Factory Method, by passing a topology context identifier to a single factory function that uses complex conditional logic to return the correct polymorphic instance. \\
\hline
Structural (SPA) & A global financial institution is redesigning its core ledger system utilizing a Hexagonal Architecture (Ports and Adapters) structural pattern to decouple the core domain logic from infrastructure concerns. The system must support high-frequency trading inputs via specialized messaging queues and expose standard RESTful endpoints for regulatory reporting. Concurrently, the domain must persist highly sensitive transaction data to a distributed SQL cluster while pushing audit events to an immutable append-only ledger. The architecture strictly mandates that all business rules reside within the core domain, which exposes Driving Ports (interfaces for incoming interactions) and Driven Ports (interfaces for outgoing interactions). The development team is attempting to optimize the structural implementation by merging the boundaries between the application services and the infrastructure adapters. During a code review, the principal architect notices a structural anomaly: a developer has designed an infrastructure component (an Adapter) that directly implements an interface defined by another infrastructure component, bypassing the domain's Driven Port entirely to improve write latency to the audit ledger. Furthermore, a Driving Adapter is injecting a Driven Adapter directly. In standard Hexagonal Architecture, the fundamental invariant is the dependency inversion principle, ensuring that all dependencies point inward toward the core domain. The core domain is entirely agnostic of the outside world, relying on the dependency injection framework to supply concrete implementations of Driven Ports at runtime. Driving Adapters, such as the REST controllers and Kafka consumers, map external requests into the core domain's vocabulary by invoking the Driving Ports. By creating a direct structural dependency where a Driving Adapter bypasses the core and directly invokes a Driven Adapter, the system violates the primary boundary condition. From the perspective of the strict Hexagonal Architecture structural pattern, what is the most critical architectural failure mode introduced by this topology? Options: (a) The direct interaction structurally binds the bounded context's ubiquitous language to the infrastructure tier, forcing the core domain to implement Anti-Corruption Layers (ACLs) internally rather than at the architectural boundaries where they belong. (b) By allowing the infrastructure to bypass the domain, the system creates a cyclic dependency graph within the dependency injection container, inevitably leading to runtime resolution failures when the Application Context attempts to initialize the messaging queue. (c) \textbf{The direct coupling between the Driving Adapter and the Driven Adapter establishes a side-channel structural dependency that completely circumvents the domain's invariant enforcement, rendering the domain model an anemic pass-through for those specific flows and destroying the pattern's fundamental testability isolation.} (d) The topology inherently violates the symmetric nature of the ports, as Hexagonal Architecture mandates that for every Driving Port there must be an exact one-to-one mapped Driven Port to ensure balanced interface segregation and strictly symmetric data flow. \\
\hline
Behavioral (BP) & In a high-frequency trading system employing the Reactor pattern for handling market data feeds, the synchronous event demultiplexer (e.g., epoll) notifies the application when a socket becomes readable. However, under extreme network burst conditions, the demultiplexer may indicate readability while the underlying receive buffer contains only partial messages due to TCP segmentation. Which of the following statements accurately describes a subtle, often overlooked behavioral consequence specific to the Reactor pattern in this scenario, and how it impacts message boundary reconstruction when using a stateful protocol decoder? Options: (a) The Reactor pattern guarantees that each readiness notification corresponds to a complete application-layer message, thereby eliminating the need for explicit buffering; any observed fragmentation is solely due to misconfiguration of the demultiplexer's edge-triggered mode and can be resolved by switching to level-triggered notifications. (b) The Reactor pattern shifts responsibility for message framing to the operating system's TCP stack, which performs automatic defragmentation when the socket is marked readable, meaning that the application can safely rely on a single read per readiness event without additional buffering, and any observed fragmentation is indicative of a broken NIC driver. (c) \textbf{Because the Reactor pattern delivers readiness events synchronously within the event loop's thread, the application may mistakenly assume that a single read operation will consume all available data, leading to incomplete message decoding if the decoder does not maintain a persistent receive buffer across successive readiness notifications, a failure mode exacerbated when the demultiplexer reports readability for bytes that do not yet constitute a full message.} (d) In the Reactor pattern, the demultiplexer automatically aggregates incoming bytes into complete messages before notifying the application, so any need for manual buffering indicates a design flaw in the protocol decoder rather than a limitation of the pattern itself; this buffering occurs transparently within the kernel's socket receive side scaling (RSS) mechanism. \\
\hline
Quality Attributes (QA) & As a principal architect for a global telecommunications provider, you are tasked with designing a subscriber management system distributed across five geographically distinct data centers. The primary architectural driver is extreme Availability (targeting 99.999\% uptime) to ensure that users can always authenticate and access network services, even during catastrophic network partitions or regional backbone failures. According to the CAP theorem and the PACELC extension, you must make explicit trade-offs between Availability and Consistency during a network partition, and between Latency and Consistency during normal operations. The subscriber data model is relatively static but includes a highly dynamic `current\_data\_usage` counter that dictates throttling policies. If a network partition isolates the Asian and European data centers, both regions must continue to accept authentication requests and log data usage to prevent service outages (prioritizing Availability over strong Consistency). However, once the partition resolves, the system must accurately reconcile the distributed `current\_data\_usage` counters without losing any recorded usage data, while avoiding complex manual intervention. Which distributed systems tactic provides the most robust mechanism to ensure high availability during the partition while guaranteeing mathematically correct, automated reconciliation of the usage counters upon recovery? Options: (a) Enforce distributed locking using a highly available centralized coordination service (e.g., Apache ZooKeeper) to ensure mutually exclusive access to the subscriber record, halting usage tracking in isolated regions until the global lock can be re-established. (b) Implement a strict quorum-based consensus algorithm (like Paxos or Raft) across all data centers, ensuring that any update to the usage counter is acknowledged by a majority of nodes before proceeding, thereby preventing split-brain scenarios. (c) \textbf{Utilize State-based Conflict-free Replicated Data Types (CvRDTs), specifically a Grow-only Counter (G-Counter) or Positive-Negative Counter (PN-Counter), allowing isolated regions to autonomously increment local state and deterministically merge states post-partition.} (d) Deploy a primary-replica architecture with asynchronous log shipping, where one data center acts as the immutable primary; during a partition, secondary regions buffer updates locally and apply a Last-Write-Wins (LWW) conflict resolution strategy upon reconnection. \\
\hline
Architectural Solutions (ASO) & In a distributed enterprise architecture utilizing CQRS and Event Sourcing to manage high-throughput financial transactions, the system relies on an append-only event store (Kafka) and a heavily indexed relational read model (PostgreSQL). During a severe disaster recovery drill, it is discovered that rebuilding the read model projections from the foundational event stream takes an unacceptable 14 hours, violating the strict 2-hour Recovery Time Objective (RTO). The stream contains over 8 billion granular domain events. The business mandates that the read model must retain exact transaction history without arbitrary truncation, and the current projection logic involves complex synchronous aggregations across multiple stream partitions. To drastically reduce the read model projection rebuild time while maintaining absolute data integrity, deterministic replayability, and adherence to the 2-hour RTO constraint, which architectural intervention is the most effective? Options: (a) Discard the persistent PostgreSQL read model entirely and route all query requests directly to the Kafka event stream utilizing KSQL for real-time, on-the-fly projection calculations during every read invocation. (b) Implement a nightly batch job that archives events older than 30 days to cold storage, permanently deleting them from the primary Kafka topics to reduce the total number of events requiring replay during a disaster recovery scenario. (c) \textbf{Implement rolling state snapshots of the aggregate roots at dynamically calculated intervals based on state mutation entropy, coupled with parallelized projection consumers that independently reconstruct localized read models segmented by bounded context and aggregate ID.} (d) Transition the underlying event store from an append-only Kafka topic to a mutable Cassandra column family, allowing in-place updates of the events to pre-calculate the projection state directly within the event store. \\
\hline
Scalable Practices (SPR) & A marketplace platform serves product detail pages from a three-layer cache: edge CDN fragments, a regional object cache, and an in-process cache inside the read service. The rendered page is not a simple key-value lookup. Each page includes a base product record, dynamic price, promotion eligibility, inventory availability by fulfillment region, and a fraud-derived badge that is itself computed from a separate event stream. Different upstream sources publish changes with different clocks and different durability guarantees. The platform cannot afford synchronous fan-out on read, because peak traffic exceeds ten million requests per minute and the origin databases are already near safe utilization during flash sales. The current implementation relies on TTLs plus best-effort invalidation messages, but the system repeatedly serves combinations that never existed in the source of truth, such as a badge computed from version N of a product policy combined with price version N+2 and inventory version N-3. Product managers now require a stronger rule: every cache hit must correspond either to a causally valid composition or be rejected and recomputed, while preserving high cache hit ratios and avoiding global coordination. Writes are frequent for price and stock, infrequent for product metadata, and bursty for promotion rules. Some updates arrive out of order, duplicate invalidation messages are common, and occasionally the fraud stream lags by minutes. Engineers debate whether shorter TTLs, write-through, or a central invalidation coordinator are sufficient. The real challenge is not simple staleness, but preventing semantically impossible mixtures of independently changing dependencies when cached artifacts are materialized at different layers and times. Which architecture practice is the most robust way to preserve correctness at scale without collapsing cache efficiency or introducing a global serialization point? Options: (a) Move all three cache layers to synchronized short TTLs and accept bounded inconsistency because sufficiently low expirations make invalid combinations statistically negligible during bursts. (b) Convert the regional cache into a write-through store so that any update path immediately recomputes every downstream fragment and pushes replacements to the edge and process caches before acknowledging the write. (c) \textbf{Attach dependency version metadata or epochs to every derived cache artifact, propagate source revision information through materialization, and reject or lazily refresh artifacts whose dependency set is no longer causally compatible with the latest known upstream revisions.} (d) Keep the current TTL design but guarantee ordered invalidation delivery per entity key with a durable message bus and a per-key mutex in the read service to prevent overlapping recomputations. \\
\hline
Architectural Styles (AST) & Consider a highly concurrent distributed application utilizing a Space-Based Architecture (SBA) to mitigate database bottleneck issues during unpredictable, massive thundering-herd traffic spikes. The system is deployed across a heterogeneous cluster with a virtualized middleware layer managing the tuple space. The processing unit consists of an application logic module, an in-memory data grid (IMDG), and an asynchronous persistence engine. Under nominal load, the caching mesh actively replicates state changes across all active processing units using an active-active, synchronous replication strategy to ensure strict serializability. However, during a sustained 400\% traffic spike, the system begins experiencing severe distributed deadlocks and cascading transaction rollbacks. To resolve this without abandoning the Space-Based paradigm or compromising eventual persistence, you analyze the replication topology and the concurrency control mechanism. Which of the following architectural modifications represents the mathematically and theoretically optimal approach to resolving the deadlock cascade while maintaining the highest possible throughput and preserving the core principles of the SBA data grid? Options: (a) Implement a localized optimistic concurrency control (OCC) mechanism within each processing unit combined with an asynchronous, fire-and-forget delta-replication protocol, resolving conflicts using a Last-Write-Wins (LWW) strategy based on global NTP timestamps. (b) Introduce a centralized distributed locking manager (e.g., ZooKeeper or etcd) alongside the tuple space to enforce strict, global pessimistic locking on all mutable tuples before any processing unit can initiate an atomic mutation operation. (c) \textbf{Migrate the processing units to utilize a partitioned data grid topology with asynchronous replication for cross-partition redundancy, routing requests via a space router utilizing dynamic consistent hashing to ensure any specific data tuple is exclusively mutated by a single designated processing unit.} (d) Transition the IMDG from synchronous active-active replication to a synchronous primary-backup topology, funneling all write operations for a given tuple through a deterministic hash-ring coordinator, thereby eliminating the multi-master distributed lock contention at the cost of localized latency. \\
\hline
Quantum Computing (QC) & In a software architecture designed to support dynamic quantum circuits featuring mid-circuit measurements and classical feed-forward operations, strict deterministic execution bounds are paramount. The target platform is an advanced trapped-ion quantum computer utilizing a micro-architectural design where classical logic must conditionally trigger subsequent quantum operations within the exact coherence lifespan of the unmeasured idling qubits. The core architectural challenge resides in the classical control software stack: the extraction of the measurement syndrome, its subsequent classification via a hardware-accelerated state discriminator, the execution of the Boolean logic evaluation, and the final dispatch of the conditional sequence trigger to the arbitrary waveform generators (AWGs) must all complete within a hard real-time latency budget of precisely 5.0 microseconds. Your current software stack employs a layered design where the measurement readout is buffered via Direct Memory Access (DMA) to a highly optimized C++ classical processor node running a preemptive real-time operating system (RTOS), which subsequently queries a pre-compiled directed acyclic graph (DAG) of feed-forward rules. Extensive profiling of this classical software pipeline demonstrates unpredictable jitter exceeding 12.0 microseconds during peak execution phases. This jitter is decisively traced to cache miss penalties within the CPU when traversing the massive memory-mapped DAG for complex algorithmic branching logic. To completely eliminate this classical software bottleneck and guarantee deterministic sub-microsecond feed-forward execution, a radical architectural paradigm shift is mandatory. What software-hardware co-design architecture must be implemented to replace the existing CPU-RTOS evaluation pipeline for evaluating the feed-forward DAG? Options: (a) Architecting a hybrid neural-network classifier within the measurement pipeline capable of predicting the classical Boolean branch prior to measurement completion, thereby overlapping computation with QPU idle time. (b) Refactoring the C++ evaluation engine to utilize highly vectorized Single Instruction Multiple Data (SIMD) instructions to compute all possible branches of the DAG simultaneously, minimizing sequential memory accesses. (c) \textbf{Translating the feed-forward DAG directly into a deterministic finite automaton (DFA) deployed as synthesized logic on an integrated Field Programmable Gate Array (FPGA) co-processor directly adjacent to the measurement digitizers.} (d) Implementing a specialized memory management unit (MMU) within the CPU architecture to pin the feed-forward DAG explicitly into the L1 data cache, disabling all external interrupts during the measurement phase. \\
\end{longtable}

\end{document}